\pdfoutput=1
\documentclass[aps,pra,twocolumn,10pt,superscriptaddress,longbibliography,floatfix]{revtex4-2}
\usepackage{graphicx}
\usepackage{amsmath,amssymb}
\usepackage{bm}
\usepackage[colorlinks=true,linkcolor=blue,citecolor=blue,urlcolor=blue]{hyperref}
\hypersetup{pdftitle={Aharonov-Bohm blockade of complete reversal in coupled Rabi cavities},pdfauthor={Shu He and Liwei Duan}}
\graphicspath{{figs/}}
\newcommand{\ManuscriptAuthors}{%
  \author{Shu He}%
  \email{heshu1987@foxmail.com}%
  \affiliation{Department of Physics and Electronic Engineering, Sichuan Normal University, Chengdu 610066, China}%
  \author{Liwei Duan}%
  \affiliation{Department of Physics, Zhejiang Normal University, Jinhua 321004, China}%
}
\newcommand{\FundingAcknowledgment}{This work was supported by the National Natural Science Foundation of China (Grant Nos. 11804240 and 12305032) and the Natural Science Foundation of Zhejiang Province (Grant No. LQ23A050003).}
\newcommand{\AIStatement}{OpenAI Codex was used to assist with language editing, checking analytical derivations and numerical convergence, and preparing the data archive.}

\newcommand{\tabtop}{}
\newcommand{\tabbot}{}
\newcommand{\cpp}{{+}{+}}
\newcommand{\cmm}{{-}{-}}
\newcommand{\cpm}{{+}{-}}
\newcommand{\cmp}{{-}{+}}
\newcommand{\ket}[1]{|#1\rangle}
\newcommand{\bra}[1]{\langle#1|}
\newcommand{\kpp}{\ket{\cpp}}
\newcommand{\kmm}{\ket{\cmm}}
\newcommand{\kpm}{\ket{\cpm}}
\newcommand{\kmp}{\ket{\cmp}}
\newcommand{\onep}{\ket{1_+}}
\newcommand{\onem}{\ket{1_-}}
\newcommand{\Jgone}{J_{\mathrm{g}1}}
\newcommand{\Jgtwo}{J_{\mathrm{g}2}}
\newcommand{\Jpi}{J_{\pi}}
\newcommand{\Jtwopi}{J_{2\pi}}
\newcommand{\Jmin}{J_{\min}}

\newcommand{\GPhi}{\Gamma_{\Phi}}
\newcommand{\Ac}{\mathcal{A}_{\mathrm{c}}}
\newcommand{\Ae}{\mathcal{A}_{\mathrm{e}}}

\newcommand{\qloc}{q_{\mathrm{loc}}}

\newcommand{\dd}{\mathrm{d}}
\newcommand{\Hc}{\mathrm{H.c.}}

\begin{document}

\title{Aharonov--Bohm blockade of complete reversal in coupled Rabi cavities}

\ManuscriptAuthors

\date{\today}

\begin{abstract}
When a common bias reverses two bistable fields, the system can pass through either of two single-reversal configurations on its way to the favored state. We show that interference between these two orders of reversal can block the final step. In a dimer of coupled quantum Rabi cavities, the four field configurations form a loop, and a complex photon-hopping amplitude threads it with an Aharonov--Bohm flux. Near half a flux quantum, the two paths to complete reversal interfere destructively: in microscopic simulations, a slow uniform sweep reverses one field with probability 0.998, although the final bias favors reversing both. A four-state model calibrated at zero bias predicts how the window of this Aharonov--Bohm blockade narrows with sweep time. The blocked state remains coherent. A differential pulse followed by a return sweep converts the relative phase of the two single-reversal configurations into a fringe in the joint field signs, with contrast 0.95. In a trimer, a phase gradient imprinted between sweeps determines whether the next sweep adds or removes a reversed field: reversing the second of two gradients changes the most probable outcome from three reversed fields to one.

\end{abstract}

\maketitle

\section{Introduction}
\label{sec:intro}

A slow bias sweep reverses a bistable system by carrying its ground state from one minimum to the other. The scenario is ubiquitous, from magnets in a field ramp to qubits swept through an avoided crossing~\cite{Landau1932,Zener1932,Stueckelberg1932,Majorana1932,Shevchenko2010}, and sweeps of many coupled bistable units underlie quantum annealing~\cite{Kadowaki1998,Albash2018}, including proposals based on networks of nonlinear oscillators~\cite{Goto2016,Puri2017}. Even two identical units driven by a common bias pose a question that a single unit cannot. The final bias favors reversing both, but either unit can reverse first, and the two intermediate configurations have the same bias energy. The system thus has two routes to the favored configuration. Can interference between these two orders of reversal prevent the system from reaching it?

Quantum interference can suppress transitions that energetics alone would allow. A geometric phase can quench tunneling between opposite spin orientations~\cite{Loss1992,vonDelft1992,Garg1993,Wernsdorfer1999}, path interference shapes transition probabilities in multistate Landau--Zener models~\cite{Brundobler1993,Ostrovsky1997,Sinitsyn2015}, and in a $\gamma$-magnet a suitable sweep rate can select one spin to flip while suppressing flips of the others~\cite{Chernyak2019}. In real space, the same principle confines a particle to an Aharonov--Bohm cage when a lattice plaquette encloses half a flux quantum~\cite{Vidal1998,Mukherjee2018}, and caging also occurs in the Fock space of two photons~\cite{Martinez2023}. In the problem posed above, the interfering paths are of a different kind. They are the two orders in which distinct cavity fields reverse, and their relative phase is set by the hopping amplitude on a single physical bond.

Coupled quantum Rabi cavities supply both ingredients that this question requires: bistable fields and a controllable hopping phase. The quantum Rabi model~\cite{Rabi1937,Braak2011,FornDiaz2019} undergoes a superradiant transition in the limit of an infinite qubit-to-cavity frequency ratio~\cite{Hwang2015}, the single-mode counterpart of the Dicke transition~\cite{Hepp1973,Baumann2010}. At large but finite ratios, its lowest doublet defines two states localized at opposite field displacements, which play the role of the two minima of a bistable field. Trapped ions have been used to simulate the Rabi model~\cite{Lv2018} and, following a theoretical proposal~\cite{Puebla2017}, to probe its superradiant transition~\cite{Cai2021}, and coupled Rabi sites support collective and chiral phases~\cite{Schiro2012,Mei2022,Zhang2021,FallasPadilla2022,Zhao2022,He2026}. Complex hopping amplitudes, for their part, can be engineered through artificial gauge fields for neutral atoms~\cite{Dalibard2011} and photons~\cite{Ozawa2019,Koch2010,Roushan2017}, and trapped-ion chains support phonon hopping~\cite{Debnath2018} and spin couplings with tunable phases~\cite{Manovitz2020,Shapira2023}. We ask what such a phase does to the reversal of two bistable fields under a uniform bias sweep.

We show that the four field configurations of a Rabi dimer form a loop in a synthetic dimension~\cite{Celi2014}, threaded by an Aharonov--Bohm flux~\cite{Aharonov1959} that a single complex hopping amplitude controls. Near half a flux quantum, destructive interference between the two orders of reversal blocks complete reversal, even when a slow sweep reverses one field with near-unit probability. We refer to this effect as Aharonov--Bohm blockade. A four-state model, calibrated from zero-bias energies and matrix elements of the bias operator, predicts how the window of the blockade narrows with sweep time. The blockade does not erase phase information: a local pulse followed by a return sweep reveals the coherence between the two single-reversal configurations. In a trimer, the same kind of phase becomes a means of control, and the direction of a local phase gradient determines whether the next sweep adds or removes a reversed field.

Section~\ref{sec:model} introduces the model and the sweep protocol and derives the configuration ring from a symmetry of the zero-bias Hamiltonian. Section~\ref{sec:results} establishes the blockade in the dimer, tests its coherence, and turns to stepwise reversal in a trimer. Section~\ref{sec:conclusion} summarizes the results, and Appendixes~\ref{app:ring}--\ref{app:pulse} collect the microscopic calibration, the finite-sweep theory, and the pulse protocols.

\section{Model and sweep protocol}
\label{sec:model}

We consider two quantum Rabi cavities coupled by photon hopping with a complex amplitude,
\begin{align}
H ={}& \sum_{j=1,2}\Big[a_j^\dagger a_j+\frac{\eta}{2}\sigma_j^z+\lambda\sqrt{\eta}\,X_j\sigma_j^x+\varepsilon(t)X_j\Big]\nonumber\\
&+J\big(e^{i\theta}a_1^\dagger a_2+\Hc\big),
\label{eq:H}
\end{align}
where $a_j$ annihilates a photon in cavity $j$, $X_j=a_j+a_j^\dagger$, $\sigma_j^{x,z}$ are Pauli matrices of the two-level system in cavity $j$, $\eta$ is the ratio of the qubit and cavity frequencies, and $\lambda$ is the dimensionless light--matter coupling. We set $\hbar=1$ and the cavity frequency to unity, and use $\eta=12$, $\lambda=0.7$, and $\theta=\pi/2$ unless stated otherwise. Only the uniform bias $\varepsilon(t)$ varies during a sweep. The bond phase $\theta$ is held fixed and, as shown below, controls the flux $\Phi$ through the configuration loop. In the trimer, the hopping is oriented cyclically around the triangle.

\begin{figure*}[t]
\centering
\includegraphics[width=\textwidth]{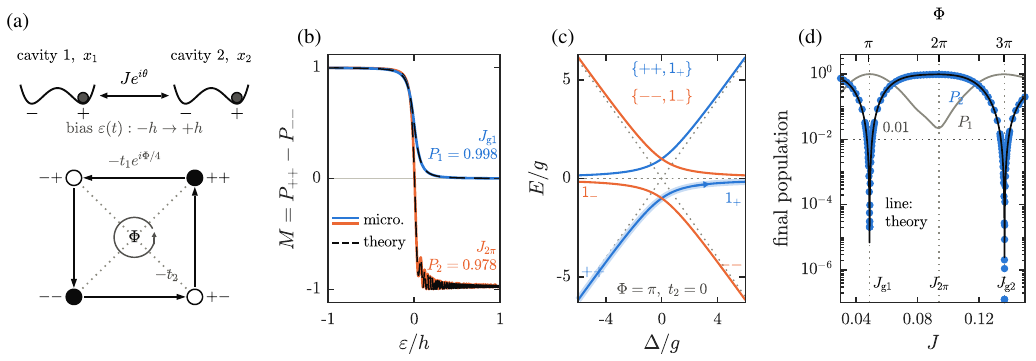}
\caption{Interference blocks complete reversal near half a flux quantum. (a)~Two Rabi cavities with bistable fields, coupled by photon hopping $Je^{i\theta}$, and the four configurations in the plane of the two fields. Arrows mark single flips with amplitude $-t_1e^{i\Phi/4}$, and dotted diagonals mark direct double flips with amplitude $-t_2$. (b)~Magnetization during the sweep at $\Jtwopi$ and $\Jgone$ for $T=2.4\times10^4$ (solid, microscopic model; dashed, two independent cavities at $\Jtwopi$ and the block $\{\kpp,\onep\}$ at $\Jgone$, Sec.~\ref{sec:blockade}). (c)~Levels of the two blocks of Eq.~\eqref{eq:blocks} versus the bias energy; the shaded curve is the path of a sweep that starts in $\kpp$, and dotted lines are the configuration energies. (d)~Final $P_2$ (symbols, microscopic model; line, Eq.~\eqref{eq:LZ} with the first-order refinement of Appendix~\ref{app:lz}) and microscopic $P_1$ (gray) versus $J$ for $T=2.4\times10^4$; the top axis gives the flux. All panels use $\eta=12$, $\lambda=0.7$, $\theta=\pi/2$, and $h=0.0125$.}
\label{fig:dimer}
\end{figure*}

An isolated cavity becomes superradiant for $\lambda>1/2$ as $\eta\to\infty$~\cite{Hwang2015}, and its field is then bistable. Eliminating the two-level system adiabatically yields a double-well potential for the field quadrature $x_j=X_j/\sqrt2$, with a barrier of height 1.41 for our parameters; tunneling between the two wells is suppressed exponentially in $\eta$ (Appendix~\ref{app:ring}). We denote the dressed states localized in the two wells of a cavity by $\ket{\pm}$ and the four configurations of the dimer by $\kpp$, $\kmp$, $\kmm$, and $\kpm$, the first sign referring to cavity 1 [Fig.~\ref{fig:dimer}(a)]. The bias shifts each well in proportion to its mean field, so the configuration with signs $s_j=\pm1$ acquires the bias energy $q\varepsilon(s_1+s_2)$. The projected bias coupling $q$ is approximately the magnitude of the field displacement $\langle X_j\rangle$ in a localized state; at the first zero-bias ground-state crossing, $\Jgone$, we find $q=4.04$. Reversing one field therefore releases the energy $\Delta(t)=2q\varepsilon(t)$, and a positive bias favors $\kmm$. The single-reversal configurations $\kmp$ and $\kpm$, by contrast, have the same bias energy throughout the sweep, which keeps the two orders of reversal on an equal footing.

The dimer is prepared in its ground state at $\varepsilon=-h$, which lies close to $\kpp$, and the bias is swept linearly to $+h$, $\varepsilon(t)=-h+2ht/T$, with $h=0.0125$ unless stated otherwise. At $t=T$ we project the state onto four orthonormal dressed configurations constructed from the instantaneous eigenstates (Appendix~\ref{app:ring}). This projection yields the probability $P_N$ that exactly $N$ fields have reversed and the magnetization $M=P_{\cpp}-P_{\cmm}$. A second readout bypasses the dressed configurations altogether: the probability $\Pr(x_1<0,\,x_2<0)$ that both fields are negative at the end of the protocol requires no effective model and corresponds directly to a measurement of the two field quadratures.

Tunneling couples the four configurations, and the form of this coupling at zero bias follows from a symmetry of Eq.~\eqref{eq:H}. At zero bias and $\theta=\pi/2$, Eq.~\eqref{eq:H} commutes with $C_4=\Pi_1S_{12}$, where $\Pi_1=-\sigma_1^z(-1)^{a_1^\dagger a_1}$ is the parity of cavity 1 and $S_{12}$ exchanges the two cavities; a finite uniform bias or a real hopping component $J\cos\theta\ne0$ breaks this symmetry. The operator $C_4$ rotates the plane of the two fields by $\pi/2$ and permutes the configurations cyclically, $\kpp\to\kmp\to\kmm\to\kpm\to\kpp$, so the zero-bias Hamiltonian restricted to these configurations is a circulant matrix. Adding the leading diagonal bias gives the ring Hamiltonian
\begin{align}
H_{\mathrm R}={}&-t_1e^{i\Phi/4}\big(\ket{\cmp}\bra{\cpp}+\ket{\cmm}\bra{\cmp}\nonumber\\
&\qquad\qquad+\ket{\cpm}\bra{\cmm}+\ket{\cpp}\bra{\cpm}\big)\nonumber\\
&-t_2\big(\ket{\cmm}\bra{\cpp}+\ket{\cpm}\bra{\cmp}\big)+\Hc\nonumber\\
&+\Delta(t)\big(\ket{\cpp}\bra{\cpp}-\ket{\cmm}\bra{\cmm}\big),
\label{eq:ring}
\end{align}
where $t_1>0$ sets the single-flip magnitude, $-t_2$ is the signed direct double-flip matrix element, and $\Phi$ is the flux through the loop [Fig.~\ref{fig:dimer}(a)]. The bias term is the bias energy introduced above, measured from the single-reversal configurations. At zero bias the ring energies, measured from their mean and labeled by the eigenvalues $i^m$ of $C_4$, are
\begin{equation}
E_m=-2t_1\cos\Big(\frac{\pi m}{2}+\frac{\Phi}{4}\Big)-t_2(-1)^m.
\label{eq:Em}
\end{equation}
Matching Eq.~\eqref{eq:Em} to the lowest zero-bias quartet of Eq.~\eqref{eq:H} determines $t_1$, $t_2$, and $\Phi$, and the matrix elements of $X_1+X_2$ within the same quartet determine $q$ (Appendix~\ref{app:ring}). All parameters of the ring are thus fixed at zero bias. At the hopping strength $\Jgone=0.0488$ used below, for example, $t_1=2.97\times10^{-3}$ and $t_2=4.87\times10^{-5}$; Table~\ref{tab:points} lists the parameters at the four reference points of the dimer.

The flux is more than a calibration parameter; it has a direct geometric meaning. In terms of the quadratures $x_j$ and their conjugate momenta $p_j$, the hopping term contains $-J\sin\theta\,(x_1p_2-x_2p_1)$, the generator of rotations in the plane of the two fields. Completing the square in the photon kinetic energy yields the perpendicular synthetic magnetic field
\begin{equation}
B=\frac{2J\sin\theta}{1-J^2\cos^2\theta}.
\label{eq:B}
\end{equation}
The flux through the loop can then be estimated as $BA$, where $A$ is the area enclosed by the four dressed-state centers. At $\Jgone$ this gives $A=32.6$ and reproduces the spectral flux to $0.13\%$ (Appendix~\ref{app:ring}). Increasing $J$ from 0.03 to 0.15 raises $\Phi$ through $\pi$, $2\pi$, and $3\pi$ [top axis of Fig.~\ref{fig:dimer}(d)], so the hopping strength alone tunes the interference between the two orders of reversal.

\section{Interference and control of field reversal}
\label{sec:results}

\subsection{Blockade of complete reversal}
\label{sec:blockade}

Figure~\ref{fig:dimer}(b) contrasts two sweeps of duration $T=2.4\times10^4$ that differ only in the hopping strength. At $\Jtwopi=0.0941$, where $\Phi=2\pi$, the sweep follows the bias: both fields reverse with probability $P_2=0.978$, and the magnetization swings from nearly $+1$ to $-0.978$. At $\Jgone$, where the two lowest zero-bias levels cross close to $\Phi=\pi$, the same sweep stops halfway. The final magnetization is only $1.65\times10^{-3}$; one field reverses with probability $P_1=0.998$, whereas $P_2=3.07\times10^{-5}$. The direct readout gives $\Pr(x_1<0,\,x_2<0)=1.15\times10^{-3}$, a larger but still small value, because the dressed single-reversal configurations are not confined to a single quadrant of the field plane. Neither energetics nor frozen dynamics can account for this outcome. At the end of the sweep $\kmm$ lies $2qh=0.101$ below the single-reversal configurations, more than 30 times $t_1$, and the first reversal occurs with near certainty.

The explanation lies in the two orders of reversal [Fig.~\ref{fig:dimer}(a)]. Starting from $\kpp$, a single flip does not select one cavity; it produces the single-reversal state
\begin{equation}
\onep=\tfrac{1}{\sqrt2}\big(e^{i\Phi/4}\kmp+e^{-i\Phi/4}\kpm\big)
\label{eq:one}
\end{equation}
with amplitude $-g$, where $g=\sqrt2\,t_1$. The final configuration $\kmm$, in turn, is reached by a single flip only from $\onem=(e^{-i\Phi/4}\kmp+e^{i\Phi/4}\kpm)/\sqrt2$. Whether the second step can proceed therefore depends on the overlap
\begin{equation}
\langle1_-|1_+\rangle=\cos(\Phi/2),
\label{eq:overlap}
\end{equation}
which is proportional to the sum of the amplitudes of the two paths $\cpp\to\cmp\to\cmm$ and $\cpp\to\cpm\to\cmm$, carrying the phases $\pm\Phi/2$. At $\Phi=2\pi$ a gauge transformation makes all single-flip amplitudes real and equal, and for $t_2=0$ Eq.~\eqref{eq:ring} describes two independent cavities. At $\Phi=\pi$ the two paths cancel and the two states are orthogonal; for $t_2=0$, Eq.~\eqref{eq:ring} then separates into the two blocks
\begin{equation}
H_\pm=\begin{pmatrix}\pm\Delta&-g\\-g&0\end{pmatrix}
\label{eq:blocks}
\end{equation}
in the bases $\{\kpp,\onep\}$ and $\{\kmm,\onem\}$ [Fig.~\ref{fig:dimer}(c)]. The first flip from $\kpp$ remains allowed, but the state it reaches has no matrix element to $\kmm$, and the second flip is blocked. This is the mechanism of an Aharonov--Bohm cage~\cite{Vidal1998,Mukherjee2018}, with configurations in place of lattice sites and the two orders of reversal in place of the two paths around a plaquette, and it is the origin of the Aharonov--Bohm blockade.

Both cases reduce to Landau--Zener problems, which makes the contrast in Fig.~\ref{fig:dimer}(b) quantitative. We measure the sweep rate by $v=\dd\Delta/\dd t=4qh/T$ and define the adiabaticity parameter $\xi=g/\sqrt v$, which equals 1.45 at $\Jgone$ for $T=2.4\times10^4$. At $\Phi=2\pi$ and $t_2=0$, each field undergoes its own passage with coupling $t_1=g/\sqrt2$, and an infinite sweep reverses both with probability
\begin{equation}
P_2=\big(1-e^{-\pi\xi^2}\big)^2,
\label{eq:P2indep}
\end{equation}
which equals 0.974 at $\Jtwopi$. The exact solution of the same two-level problems over the finite sweep gives 0.978 and follows the microscopic magnetization throughout the sweep [dashed line in Fig.~\ref{fig:dimer}(b)]. At $\Phi=\pi$ and $t_2=0$, the sweep starts in $\kpp$ and never leaves the block $H_+$, whose lower level connects $\kpp$ to $\onep$ [shaded curve in Fig.~\ref{fig:dimer}(c)]. The infinite-sweep result is then
\begin{equation}
P_1=1-e^{-2\pi\xi^2},\qquad P_2=0.
\label{eq:P1block}
\end{equation}
At the finite positive endpoint, the main correction is an admixture $P_0\approx g^2/\Delta_h^2$ of $\kpp$, with $\Delta_h=2qh$; at $\Jgone$ this gives $1.73\times10^{-3}$, close to the microscopic $1.68\times10^{-3}$.

The microscopic model does not meet the two ideal conditions $\Phi=\pi$ and $t_2=0$ at the same hopping strength. At $\Jpi=0.0481$, where $\Phi=\pi$, the direct amplitude is $t_2=4.80\times10^{-5}$, and the two blocks remain coupled. At zero bias their coupling equals half the signed splitting of the two lowest levels,
\begin{equation}
\Gamma=\frac{E_0-E_3}{2}=g\sin\frac{\Phi-\pi}{4}-t_2,
\label{eq:Gamma}
\end{equation}
and it vanishes at the crossing $\Jgone$ of these levels, where the flux detuning $\Phi-\pi=0.0464$ compensates $t_2$. The ideal cancellation therefore occurs not at half flux but slightly beyond it. As the bias passes through zero, the lower levels of the two blocks cross with relative slope $v$, and the Landau--Zener formula gives the population transferred to the block that ends in $\kmm$,
\begin{equation}
P_2=1-\exp\big(-2\pi\Gamma^2/v\big).
\label{eq:LZ}
\end{equation}
Evaluated with $\Gamma$ taken from the zero-bias spectrum at each $J$, and with the bias dependence of the coupling included to first order (Appendix~\ref{app:lz}), this expression reproduces the microscopic $P_2$ across the windows near $\Jgone$ and near the corresponding crossing $\Jgtwo=0.137$ close to $\Phi=3\pi$, as well as between them, with a median relative deviation of $1.2\%$ where $P_2>0.01$ [Fig.~\ref{fig:dimer}(d)].

The blockade requires an intermediate sweep rate. The sweep must be slow enough for the first reversal, $e^{-2\pi\xi^2}\ll1$, yet fast enough to suppress transfer between the blocks, $2\pi\Gamma^2/v\ll1$. At fixed $J$ away from the crossing, a longer sweep eventually permits the second reversal, so the window in $J$ narrows as $T^{-1/2}$, consistent with the narrowing seen in the microscopic sweeps (Appendix~\ref{app:lz}). Even at $\Jgone$, where the zero-bias coupling vanishes, a small residual $P_2$ remains, because the sweep spans a finite range of bias. An amplitude acquired near the crossing can interfere with the admixture of $\kmm$ at the endpoint; the estimate in Appendix~\ref{app:lz} accounts for the scale of this residual and explains why the smallest $P_2$ need not occur exactly at $\Jgone$.

\subsection{Coherence of the single-reversal state}
\label{sec:coherence}

\begin{figure}[t]
\centering
\includegraphics[width=\columnwidth]{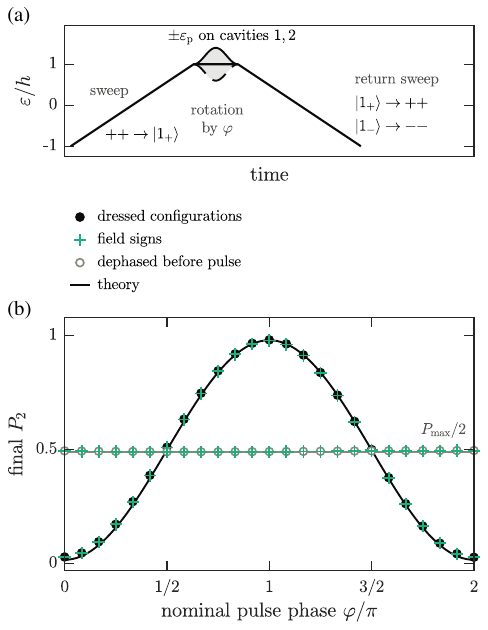}
\caption{The single-reversal state is a coherent superposition: a differential pulse and a return sweep turn it into complete reversal with a probability set by the pulse phase. (a)~Protocol: a sweep to $+h$, opposite bias pulses $\pm\varepsilon_{\mathrm p}$ with nominal phase $\varphi$ (Appendix~\ref{app:pulse}), and a sweep back to $-h$; each sweep lasts $T=10^{4}$, the pulses last $\tau=160$, and $h=0.015$. (b)~Final $P_2$ from the dressed configurations (filled circles) and from the joint signs of the two fields (crosses). Open circles and the flat row of crosses show both readouts after dephasing of the configurations before the pulse. The black line gives the finite-sweep estimate of Eq.~\eqref{eq:fringeFinite}, and the gray horizontal line marks half its maximum.}
\label{fig:phase}
\end{figure}

\begin{figure*}[t]
\centering
\includegraphics[width=\textwidth]{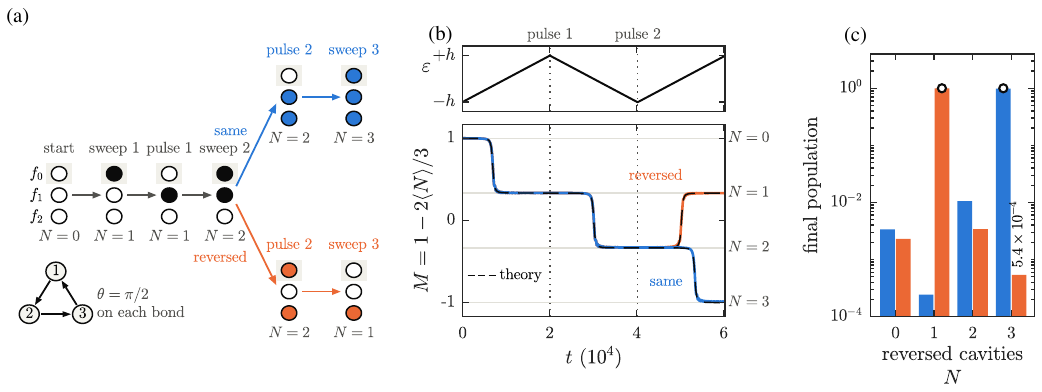}
\caption{A phase gradient controls the next reversal step in a trimer with flux $\pi$ on each face of its configuration cube. (a)~Occupations of the momentum modes $f_k$ (filled, occupied) after each step, for a second pulse identical to the first (blue) or reversed (orange); inset, the three cavities with hopping phase $\pi/2$ on each bond. (b)~Bias (top) and configuration magnetization $M$ (bottom) during three sweeps of duration $T=2\times10^{4}$ with $h=0.03$ (solid, microscopic model; dashed, adiabatic toggling of $f_0$); horizontal lines mark $N=0,\ldots,3$. (c)~Final microscopic probabilities for identical (blue, left) and reversed (orange, right) second gradients; open circles mark the unit-probability outcomes of ideal mode toggling.}
\label{fig:trimer}
\end{figure*}

Populations alone cannot distinguish the blockade from a sweep that reverses one field at random, since an incoherent mixture of $\kmp$ and $\kpm$ would yield the same $P_1$. In the ideal ring, however, the sweep prepares $\onep$, an equal-weight coherent superposition of two configurations with the same net displacement, whose field part has the structure of an entangled coherent state~\cite{Sanders1992,Joo2011,Wang2016}. For the single sweep in Fig.~\ref{fig:dimer}(b), the microscopic populations are $P_{+-}=0.4657$ and $P_{-+}=0.5326$, giving $P_1=0.9983$. It remains to show that the relative phase between them survives the sweep and can be read out.

To probe this phase, we extend the protocol by a pulse and a return sweep [Fig.~\ref{fig:phase}(a)]. At $\Jgone$ we sweep from $-h$ to $+h$, apply opposite local biases for $\tau=160$, and sweep back; each sweep has $h=0.015$, $T=10^4$, and $\xi=0.854$. The pulses shift $\kmp$ and $\kpm$ in opposite directions and imprint a relative phase between them, which we characterize by the nominal phase $\varphi=4\qloc\int\varepsilon_{\mathrm p}\,\dd t$, the value for uncoupled cavities with bias coupling $\qloc=4.02$ (Appendix~\ref{app:pulse}); the ring gives the imprinted phase $\varphi'=(q/\qloc)\varphi$. At ideal half flux, this phase rotates $\onep$ into $\cos(\varphi'/2)\onep-\sin(\varphi'/2)\onem$. The return sweep then traverses each block backward: $\onep$ returns to $\kpp$, whereas $\onem$ is carried on to $\kmm$. The result is the ideal adiabatic fringe
\begin{equation}
P_2(\varphi')=\sin^2(\varphi'/2).
\label{eq:fringe}
\end{equation}
The return sweep thus converts a relative phase into a population, as in Landau--Zener--St\"uckelberg interferometry~\cite{Shevchenko2010,Oliver2005}. At a finite sweep rate, two imperfections shape the fringe. Incomplete block passages lower its maximum to $1-2e^{-2\pi\xi^2}=0.980$, and the direct double flips, which mix $\onep$ and $\onem$ at finite bias, give it a nonzero minimum of at most 0.033 (Appendix~\ref{app:pulse}). The microscopic fringe, which ranges from 0.0276 to 0.9812, is consistent with both estimates [Fig.~\ref{fig:phase}(b)].

The same fringe appears in the field signs themselves, without reference to the dressed configurations. Evaluated at 25 pulse phases, $\Pr(x_1<0,\,x_2<0)$ has a contrast, defined as the difference between its maximum and minimum, of 0.951, close to the dressed-configuration contrast of 0.9536. Dephasing the configurations before the pulse removes the fringe: the field-sign probability stays near half the coherent maximum, and its contrast drops to 0.006 [Fig.~\ref{fig:phase}(b)]. A local phase can thus steer the same return sweep either back to $\kpp$ or on to $\kmm$, and the loss of contrast when the two reversal paths are distinguished is the expected signature of which-path dephasing~\cite{Scully1991,Englert1996}.

\subsection{Stepwise reversal in a trimer}
\label{sec:trimer}

If a local pulse can decide whether a return sweep completes the reversal, a phase imprinted on an intermediate state should also be able to steer the next step of a longer sequence. We test this idea with three cavities on a triangle, with a hopping phase $\pi/2$ on each bond, $J=0.0475$, $\eta=12$, and $\lambda=0.7$ [Fig.~\ref{fig:trimer}(a)]. The eight field configurations now form a cube. Each face describes reversals of two fields with the third held fixed, and at this working point every face carries a flux of $\pi$. We apply three alternating uniform sweeps of duration $T=2\times10^4$ and amplitude $h=0.03$, separated by two phase-gradient pulses of duration $\tau=160$ [Fig.~\ref{fig:trimer}(b)], and compare two protocols that are identical up to the second pulse. Repeating the gradient yields three reversed fields with probability $P_3=0.986$, while the single-reversal probability is $P_1=2.43\times10^{-4}$. Reversing only the second gradient changes what the same final sweep does: it now leaves a single reversed field, with probability $P_1=0.994$, and $P_3$ falls to $5.39\times10^{-4}$ [Fig.~\ref{fig:trimer}(c)].

The mode picture of Fig.~\ref{fig:trimer}(a) accounts for this behavior. In the dimer, a flux of $\pi$ blocked the second reversal; on the faces of the cube, the same flux makes the configurations equivalent to three fermion modes. In a Jordan--Wigner representation~\cite{JordanWigner1928} with $n_j=c_j^\dagger c_j=1$ when the field in cavity $j$ is reversed, the sign of the string around a face reproduces its flux $\pi$, and after a gauge transformation the single-flip terms in the equal-amplitude limit become (Appendix~\ref{app:pulse})
\begin{equation}
-t_1\sum_{j=1}^3\big(c_j+c_j^\dagger\big)=-g_3\big(f_0+f_0^\dagger\big),\qquad g_3=\sqrt3\,t_1.
\label{eq:jw}
\end{equation}
Here $t_1$ denotes the trimer single-flip scale, and the operators $f_k=3^{-1/2}\sum_je^{-2\pi ikj/3}c_j$ ($k=0,1,2$) annihilate the three momentum modes. Because the bias and the zero-bias configuration energies depend only on the number $N=\sum_kf_k^\dagger f_k$ of reversed fields, a uniform sweep is a Landau--Zener passage that changes the occupation of the uniform mode $f_0$ and leaves the other two modes untouched. The zero-bias energies nonetheless differ between shells: the $N=0,3$ configurations lie 0.0736 above those with $N=1,2$, which shifts the $0\leftrightarrow1$ and $2\leftrightarrow3$ resonances to $\varepsilon/h\simeq-0.305$ and $+0.305$, respectively, while the $1\leftrightarrow2$ resonance remains at zero [Fig.~\ref{fig:trimer}(b)]. In either sweep direction, an adiabatic passage fills $f_0$ if it is empty and empties it if it is occupied, adding or removing one reversed field. At $T=2\times10^4$ the first sweep reverses one field with probability 0.997, in agreement with $1-e^{-2\pi g_3^2/v_3}$, where $v_3=4\qloc h/T$.

The gradient pulses, in turn, move occupation between modes. A pulse that imprints the phase $2\pi(j-2)/3$ on configurations in which the field in cavity $j$ is reversed shifts each occupied mode from $k$ to $k+1$ modulo three (Appendix~\ref{app:pulse}). The protocol of Fig.~\ref{fig:trimer} then reads as a short sequence of mode operations. The first sweep fills $f_0$, the first pulse moves that occupation to $f_1$, and the second sweep fills $f_0$ again. Repeating the gradient shifts the occupied modes from $\{0,1\}$ to $\{1,2\}$, leaving $f_0$ empty, and the final sweep fills it to complete the reversal. Reversing the gradient instead shifts $\{0,1\}$ to $\{2,0\}$; now $f_0$ is occupied, and the same sweep empties it, leaving a single reversed field. For identical pulses, the microscopic $P_3=0.986$ is close to $0.997^3\simeq0.991$, the prediction for three successful passages.

\section{Conclusion}
\label{sec:conclusion}

We have shown that the amplitudes for the two orders in which a pair of bistable fields reverses can interfere, and that this interference can block complete reversal. In coupled Rabi cavities, the four field configurations form a loop whose flux is set by a single complex hopping amplitude. Near half a flux quantum, a slow sweep reverses only one field even though the final bias favors reversing both, and a four-state model calibrated at zero bias predicts how the window of this Aharonov--Bohm blockade narrows with sweep time.

The blocked state is not a random reversal of one field. Its relative phase survives the sweep and can be read out from the joint field signs after a return sweep, so the coherence of the blockade is accessible to field-quadrature measurements. In a trimer, a phase imprinted between sweeps determines whether the same final sweep adds or removes a reversed field. Because the blockade is set by the phase of a single bond, it may also be relevant to other arrays of bistable units swept by a common bias, such as the oscillator networks proposed for annealing~\cite{Goto2016,Puri2017}.

\begin{acknowledgments}
\FundingAcknowledgment{} \AIStatement{}
\end{acknowledgments}

\section*{Data availability}
The numerical data underlying Figs.~\ref{fig:dimer}--\ref{fig:windows} and Table~\ref{tab:points}, together with the simulation and plotting scripts, are available on Zenodo~\cite{HeDuanZenodo2026}.

\appendix
\section{Configuration ring and microscopic readout}
\label{app:ring}

At fixed $x=X/\sqrt2$, the lower eigenvalue of the two-level Hamiltonian in Eq.~\eqref{eq:H} gives the Born--Oppenheimer potential
\begin{equation}
 U(x)=\frac{x^2}{2}-\frac{\eta}{2}\sqrt{1+8\lambda^2x^2/\eta}.
 \label{eq:U}
\end{equation}
For $\lambda>1/2$ the potential has two minima at $\pm x_0$, separated by a barrier. Their positions and the barrier height are
\begin{equation*}
x_0^2=\frac{\eta(16\lambda^4-1)}{8\lambda^2},\qquad
U(0)-U(x_0)=\frac{\eta(4\lambda^2-1)^2}{16\lambda^2}.
\end{equation*}
For $\eta=12$ and $\lambda=0.7$ the barrier height is 1.41. Writing $x=\sqrt{\eta}\,u$ gives $U=\eta\bar U(u)$, so the semiclassical tunneling action grows linearly with $\eta$. The corresponding displacement estimate, $\sqrt2 x_0=4.17$, is close to $\qloc=|\bra{e}X\ket{o}|=4.02$ for the lowest isolated-cavity parity doublet, whose relative phase we choose such that $(\ket{e}\pm\ket{o})/\sqrt2$ have mean fields $\pm\qloc$.

Let $E_m$ denote the lowest energy in the sector $C_4=i^m$, measured from the mean of the quartet. Inverting Eq.~\eqref{eq:Em} gives
\begin{align}
 t_1 e^{i\Phi/4}&=\tfrac14\big[E_2-E_0+i(E_1-E_3)\big],\nonumber\\
 t_2&=\tfrac12(E_1+E_3).
 \label{eq:inversion}
\end{align}
The branch of $\Phi$ is fixed by continuity from $J=0$. For the uniform bias, let $D$ be the matrix of $X_1+X_2$ in the same quartet; matching $\operatorname{Tr}D^2$ to its ring value gives $q=(\operatorname{Tr}D^2/8)^{1/2}$. Table~\ref{tab:points} lists the resulting parameters and distinguishes the half-flux point $\Jpi$ from the first crossing $\Jgone$. The hopping strength $\Jmin(T)$ at which the final $P_2$ is smallest for a given sweep time can differ from both.

\begin{table*}[t]
\caption{Dimer calibration at $\eta=12$, $\lambda=0.7$, and $\theta=\pi/2$. $\Jpi$ and $\Jtwopi$ are defined by their flux, and at $\Jgone$ and $\Jgtwo$ the two lowest zero-bias levels cross. The last three columns refer to a forward sweep with $h=0.0125$ and $T=2.4\times10^4$.}
\label{tab:points}
\begin{ruledtabular}
\begin{tabular}{lcccccccc}
\tabtop
 & $J$ & $\Phi/\pi$ & $t_1\,(10^{-3})$ & $t_2\,(10^{-5})$ & $q$ & $\xi$ & $P_1$ & $P_2$ \\
\colrule
$\Jpi$ & 0.04814 & 1.000 & 2.978 & $4.796$ & 4.037 & 1.45 & 0.996 & $2.03\times10^{-3}$ \\
$\Jgone$ & 0.04884 & 1.015 & 2.972 & $4.874$ & 4.037 & 1.45 & 0.998 & $3.07\times10^{-5}$ \\
$\Jtwopi$ & 0.09407 & 2.000 & 2.424 & $4.264$ & 4.079 & 1.18 & 0.0220 & 0.978 \\
$\Jgtwo$ & 0.13671 & 3.008 & 1.765 & $-1.553$ & 4.143 & 0.85 & 0.985 & $8.61\times10^{-6}$ \\
\tabbot
\end{tabular}

\end{ruledtabular}
\end{table*}

Two independent checks support this calibration of the flux. First, the flux also follows from the phases of the directed matrix elements $H_{s's}=\bra{s'}H\ket{s}$. Their product around $\cpp\to\cmp\to\cmm\to\cpm\to\cpp$ is invariant under rephasing of the localized states, and its argument agrees modulo $2\pi$ with the spectral flux at $J=0.01$, 0.02, and 0.03 for $\theta=\pi/2$. Second, for the geometric estimate in Sec.~\ref{sec:model}, the quadratic photon Hamiltonian has momentum terms $p^{\mathsf T}Qp/2+p^{\mathsf T}Rx$, with $Q_{11}=Q_{22}=1$, $Q_{12}=Q_{21}=J\cos\theta$, $R_{12}=-R_{21}=J\sin\theta$, and $R_{11}=R_{22}=0$. Completing the square gives the vector potential $-Q^{-1}Rx$ and the uniform field in Eq.~\eqref{eq:B}. Taking the four centers $(\langle x_1\rangle,\langle x_2\rangle)$ of the dressed states as vertices gives $A=32.595$ at $\Jgone$, and the estimate $BA$ differs from $\Phi$ by $0.13\%$ there and by $0.65\%$ at $\Jgtwo$.

\begin{figure*}[t]
\centering
\includegraphics[width=.75\textwidth]{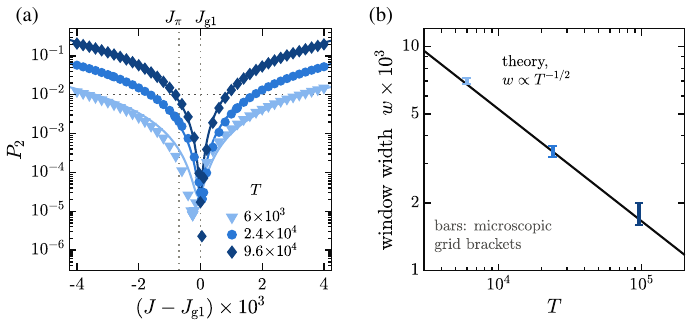}
\caption{Finite-sweep test of the blockade at $\eta=12$ and $h=0.0125$. (a)~Microscopic $P_2$ (symbols) near $\Jgone$ for three sweep durations; lines give the phase average $|\Ac|^2+|\Ae|^2$ from Eqs.~\eqref{eq:Ac} and \eqref{eq:Ae}, and the dotted horizontal line marks $P_2=0.01$. (b)~Full width of the interval below that threshold. End caps give the inner and outer widths resolved by the microscopic $J$ grid, and the line is the static prediction of Eq.~\eqref{eq:width}.}
\label{fig:windows}
\end{figure*}

For the microscopic dimer we keep 40 photon states per cavity, checked against 32, and 48 for the final static calibration. The propagation basis is built from instantaneous eigenstates. Within the space of the 64 lowest zero-bias states, we take the six lowest eigenvectors at each of five biases from $-h$ to $+h$, and the dominant left singular vectors of this combined set span a fixed basis of dimension 24. For the pulse protocol, the sampling also includes the differential bias, and the basis has 32 vectors. Each time step applies a midpoint exponential. For the trimer we keep 40 photon states and 20 dressed levels per cavity and propagate in the basis of the 192 lowest zero-bias eigenstates. Uniform sweeps conserve the cyclic momentum, whereas the gradient pulses couple the three momentum sectors; each sweep uses 15\,960 midpoint steps and each pulse 320.

The dressed configurations used in the readout are obtained by symmetric orthogonalization~\cite{Lowdin1950}. Let $L$ be the matrix of product well states and $V$ an orthonormal basis of instantaneous eigenstates. If $V^\dagger L=U\Sigma W^\dagger$, the dressed configurations are the columns of $VU W^\dagger$, and the squared singular values measure the overlap of the product states with the selected subspace before orthogonalization. For the dimer, $V$ contains the four lowest eigenstates. At the trimer endpoints $|\varepsilon|=0.03$, it contains the eight eigenstates with the largest total overlap with $L$, and the smallest squared singular value is at least 0.888. The probabilities $P_N$ retain their normalization to the full state. In the trimer we compute $M=\sum_{N=0}^3(1-2N/3)P_N$, which reduces to $1-2\langle N\rangle/3$ for unit weight in the octet, as labeled in Fig.~\ref{fig:trimer}(b). The field-sign readout uses the joint negative-half-axis projector for the two oscillator quadratures, summed over the qubit states.

The trimer results are converged well beyond the precision quoted in the text. All eight endpoint probabilities in Fig.~\ref{fig:trimer}(c) change by less than $0.01\%$ when the basis is enlarged from 144 to 192 states or the number of time steps is doubled, and they are similarly insensitive to reducing the photon cutoff to 32 states or enlarging the local dressed basis to 24 levels per cavity.

\section{Blockade window and finite-sweep leakage}
\label{app:lz}

We use the bright states at $\Phi=\pi$ as a fixed basis. In each ideal block of Eq.~\eqref{eq:blocks} the level separation is $\Omega=(\Delta^2+4g^2)^{1/2}$, and the lower levels of $H_+$ and $H_-$ have energies $(\Delta-\Omega)/2$ and $(-\Delta-\Omega)/2$, respectively; their relative slope therefore has magnitude $v$. Their weights on $\kpp$ and $\kmm$ are $(1-\Delta/\Omega)/2$ and $(1+\Delta/\Omega)/2$. Projecting a flux detuning and a direct double flip onto this pair gives the interblock matrix element
\begin{equation}
 V_{-+}(\Delta)=\GPhi-\frac{2g t_2}{\Omega},\qquad
 \GPhi=g\sin[(\Phi-\pi)/4].
 \label{eq:coupling}
\end{equation}
The factor $2g/\Omega$ arises because the direct flip couples both pairs of opposite configurations. At zero bias $V_{-+}$ reduces to $\Gamma$ of Eq.~\eqref{eq:Gamma}. Near the second crossing the same construction applies with $\GPhi=-g\sin[(\Phi-3\pi)/4]$ and $\Gamma=(E_2-E_3)/2$.

The width of the blockade window follows from the two conditions of Sec.~\ref{sec:blockade}: the sweep must be adiabatic for a single reversal and diabatic at the interblock crossing, which in the asymptotic Landau--Zener limit read $2\pi g^2/v\gg1$ and $2\pi\Gamma^2/v\ll1$. Linearizing $\Gamma$ at $J_{\mathrm g}$ gives the full width for $P_2\le P_{\mathrm c}$,
\begin{equation}
 w=\frac{2}{|\partial_J\Gamma|}\sqrt{\frac{v}{2\pi}\ln\frac{1}{1-P_{\mathrm c}}}.
 \label{eq:width}
\end{equation}
At $\Jgone$, $|\partial_J\Gamma|=0.0685$, and with $P_{\mathrm c}=0.01$ Eq.~\eqref{eq:width} gives $w=3.39\times10^{-3}$ at $T=2.4\times10^4$, inside the microscopic grid bracket $[3.2,3.6]\times10^{-3}$. For the shortest sweep, $T=6\times10^3$, the predicted width $6.77\times10^{-3}$ lies just below the microscopic bracket $[6.8,7.2]\times10^{-3}$ [Fig.~\ref{fig:windows}(b)].

The curve in Fig.~\ref{fig:dimer}(d) retains the bias dependence of Eq.~\eqref{eq:coupling}. With $t=0$ at the crossing, the lower-level energy difference is $-vt$, and integrating $V_{-+}(vt)e^{-ivt^2/2}$ over an unbounded sweep gives, to first order and up to an overall phase,
\begin{align}
 \Ac&=\sqrt{2\pi/v}\,[\GPhi-t_2 r(\xi)],\nonumber\\
 r(\xi)&=-i\sqrt{\pi/2}\,\xi\,e^{i(\xi^2+\pi/4)}H_0^{(2)}(\xi^2),
 \label{eq:Ac}
\end{align}
where $H_0^{(2)}$ is the Hankel function of the second kind and $r\to1$ as $\xi\to\infty$. Figure~\ref{fig:dimer}(d) uses $1-\exp(-|\Ac|^2)$, a continuation of the weak-transfer result that recovers Eq.~\eqref{eq:LZ} in this limit.

A second amplitude arises near the endpoint. For $\Delta_h=2qh\gg g$, the two predominantly single-reversal levels have separation $2g^2/\Delta$ and coupling $-t_2$, so their mixing angle at the positive endpoint is $\beta_h=\tfrac12\arctan(t_2\Delta_h/g^2)$. To first order in $t_2$, integrating the relative phase $2\xi^2\ln(\Delta/\Delta_h)$ gives an admixture $b\simeq\beta_h\,2i\xi^2/(1+2i\xi^2)$. Its overlap with $\kmm$, together with the direct flux contribution, yields the endpoint amplitude, up to the relative propagation phase,
\begin{equation}
 \Ae\simeq\frac{t_2}{2g}\frac{2i\xi^2}{1+2i\xi^2}
             -\frac{\GPhi}{\Delta_h}.
 \label{eq:Ae}
\end{equation}
Near the first crossing the two amplitudes interfere, and their relative dynamical phase is sensitive to the sweep duration: at $T=2.4\times10^4$ it changes by about $2\pi$ for a $1\%$ change in $T$. The lines in Fig.~\ref{fig:windows}(a) therefore show their phase-averaged probability. At $\Jgone$ the two extreme relative phases give $P_2\simeq8.0\times10^{-6}$ and $5.7\times10^{-5}$, bracketing the microscopic value $3.07\times10^{-5}$. The same phase dependence allows $\Jmin(T)$ to move to either side of $\Jgone$.

\section{Phase pulses in the dimer and trimer}
\label{app:pulse}

During each pulse the uniform bias is held at its endpoint. In the dimer, the pulse adds $+\varepsilon_{\mathrm p}$ to cavity 1 and $-\varepsilon_{\mathrm p}$ to cavity 2; for elapsed pulse time $0\le s\le\tau$ we take
\begin{equation}
 \varepsilon_{\mathrm p}(s)=\frac{\varphi}{2\qloc\tau}\sin^2(\pi s/\tau).
 \label{eq:pulse}
\end{equation}
The opposite shifts of $\kmp$ and $\kpm$ give the relative phase $\varphi'=(q/\qloc)\varphi$, with $q/\qloc=1.0039$ at $\Jgone$. At ideal half flux the pulse populates $\onep$ and $\onem$ with amplitudes $\cos(\varphi'/2)$ and $-\sin(\varphi'/2)$, and only $\onem$ connects to $\kmm$ on the return sweep.

Two effects set the fringe envelope at finite $T$. Each passage fails with probability $e^{-2\pi\xi^2}$, giving $P_{\max}\simeq1-2e^{-2\pi\xi^2}$ to leading order. In addition, the single-reversal state lags behind the endpoint rotation by $|\alpha|=|\beta_h-b|\simeq|\beta_h|/\sqrt{1+4\xi^4}$, and the two sweeps contribute lags of equal magnitude. Averaging over their relative phase gives the curve plotted against the nominal phase in Fig.~\ref{fig:phase}(b),
\begin{equation}
 P_2^{\rm est}(\varphi)=P_{\max}\big[\sin^2(\varphi/2)+2|\alpha|^2\cos^2(\varphi/2)\big].
 \label{eq:fringeFinite}
\end{equation}
Here $|\alpha|^2=8.3\times10^{-3}$. For the dephased control, we apply $\rho\mapsto\sum_a P_a\rho P_a$ at $+h$ just before the pulse, using the four zero-bias dressed-configuration projectors and their complement; the final dressed readout uses the instantaneous basis at $-h$.

In the trimer, let $n_j=0,1$ record whether the field in cavity $j$ is reversed, and let $F_j$ flip that bit. The Jordan--Wigner operators $c_j=(\prod_{l<j}(-1)^{n_l})F_j n_j$ then anticommute, and the strings give a product of hopping signs of $-1$ around each face, the same Wilson phase as the microscopic single-flip amplitudes. A rephasing of the cube vertices therefore yields Eq.~\eqref{eq:jw} for equal single-flip magnitudes. In the microscopic trimer these magnitudes differ by $2.2\%$ between adjacent shells; this shell dependence preserves the selection of $f_0$, and direct flips of two or three fields give only small corrections.

The trimer pulse adds to cavity $j$ the bias
\begin{equation*}
 \delta\varepsilon_j(s)=[\phi_j/(\qloc\tau)]\sin^2(\pi s/\tau),
 \qquad 0\le s\le\tau,
\end{equation*}
where $\phi_j=2\pi(j-2)/3$ for the forward gradient. A reversed field in cavity $j$ thus acquires the phase $2\qloc\int_0^\tau\delta\varepsilon_j(s)\,\dd s=\phi_j$ relative to the unreversed configuration, and the pulse acts as
\begin{equation*}
c_j^\dagger\mapsto e^{i\phi_j}c_j^\dagger,\qquad
f_k^\dagger\mapsto e^{-4\pi i/3}f_{k+1}^\dagger.
\end{equation*}
The opposite gradient shifts $k$ to $k-1$ instead, which produces the second outcome in Fig.~\ref{fig:trimer}(c).

\bibliography{refs}

\end{document}